\documentclass[10pt]{article}
\usepackage[a4paper,margin=0.8in]{geometry}

\DeclareUnicodeCharacter{2212}{--}
\DeclareUnicodeCharacter{03C9}{\(\omega\)}
\DeclareUnicodeCharacter{0394}{\(\Delta\)}
\DeclareUnicodeCharacter{03C3}{\(\sigma\)}

\usepackage[dvipsnames]{xcolor} %
\usepackage{graphicx}

\definecolor{msft_teal}{rgb}{0.286, 0.773, 0.694}
\definecolor{msft_light_teal}{rgb}{0.725, 0.863, 0.824}
\definecolor{msft_dark_teal}{rgb}{0.133, 0.357, 0.384}
\definecolor{msft_yellow}{rgb}{1.0,   0.725, 0.0}
\definecolor{msft_blue}{rgb}{0.0,   0.471, 0.831}
\definecolor{msft_purple}{rgb}{0.525, 0.380, 0.773}
\definecolor{msft_orange}{rgb}{1.0,   0.639, 0.545}
\definecolor{msft_white}{rgb}{1.0,   1.0,   1.0}
\definecolor{msft_black}{rgb}{0.0,   0.0,   0.0}
\definecolor{msft_grey}{rgb}{0.851, 0.851, 0.839}

\definecolor{tab10_blue}{rgb}{0.121, 0.466, 0.705}
\definecolor{tab10_orange}{rgb}{1.0,   0.498, 0.054}
\definecolor{tab10_green}{rgb}{0.172, 0.627, 0.172}
\definecolor{tab10_red}{rgb}{0.839, 0.152, 0.156}
\definecolor{tab10_purple}{rgb}{0.580, 0.403, 0.741}
\definecolor{tab10_brown}{rgb}{0.549, 0.337, 0.294}
\definecolor{tab10_pink}{rgb}{0.890, 0.466, 0.760}
\definecolor{tab10_gray}{rgb}{0.498, 0.498, 0.498}
\definecolor{tab10_olive}{rgb}{0.737, 0.741, 0.133}
\definecolor{tab10_cyan}{rgb}{0.090, 0.745, 0.811}

\def\commentType{0}
\usepackage{xspace}

\ifnum\commentType=0
  \usepackage[disable]{todonotes} %

\fi
\ifnum\commentType=1

\fi
\ifnum\commentType=2
  \usepackage{pdfcomment}

\fi
\ifnum\commentType=3
  \usepackage{todonotes} %
  \usepackage{silence}
\fi

\usepackage{fancyhdr}
\usepackage{microtype}
\usepackage[small]{titlesec}
\titleformat*{\subparagraph}{\itshape}
\usepackage[T1]{fontenc}
\usepackage[utf8]{inputenc}
\usepackage{tabularx}
\usepackage{duckuments}
\usepackage{amssymb,amsmath,amsthm,amsfonts}
\usepackage{bm}             %
\usepackage{hyperref}
\usepackage{pifont}
\usepackage{mathtools}
\usepackage{dsfont}

\usepackage{tikz}
\usepackage[version=4]{mhchem}
\usepackage{multirow}
\usepackage[exponent-product=\cdot]{siunitx}
\usepackage{booktabs}
\usepackage{xspace}
\usepackage{subcaption}
\usepackage[capitalize]{cleveref}
\usepackage{float}
\usepackage{enumitem}
\usepackage{jabbrv}
\DefineSpuriousJournalWord{The}

\usepackage[style=numeric-comp,autocite=superscript,sorting=none,natbib=true,backend=biber]{biblatex}

\let\cite\autocite%
\let\citep\autocite%
\renewcommand{\citet}[1]{\citeauthor{#1}\,\supercite{#1}}
\AtBeginBibliography{\small}

\usepackage{tocloft}
\usepackage{etoc}

\usepackage{parskip}

\usepackage[
    justification=justified,
    singlelinecheck=true,
    labelfont={bf},
    font=small
]{caption}
\begin{document}
\suppressfloats

\newcommand{\molcount}{73,040}

\begin{center}
    ~\\[0.5em]
    {\LARGE\bfseries
        Accurate Chemistry Collection:
        Coupled cluster atomization energies for broad chemical space
    }
    \vspace{1.5em}
    
    \newcommand{\maybecomma}{,}
    \renewcommand{\author}[2][]{#2${}^{\text{#1}}$\maybecomma}
\author[1,\textdagger,*]{Sebastian Ehlert}
\author[1,\textdagger,*]{Jan Hermann}
\author[1]{Thijs Vogels}
\author[1]{Victor Garcia Satorras}
\author[1]{Stephanie Lanius}\\
\author[1]{Marwin Segler}
\author[1]{Klaas J.H. Giesbertz}
\author[1]{Derk P. Kooi}
\author[2]{Kenji Takeda}
\author[1]{Chin-Wei Huang}
\author[1]{Giulia Luise}\\
\author[1]{Rianne van den Berg}
\author[1]{Paola Gori-Giorgi}\renewcommand{\maybecomma}{}
\author[1,3,*]{Amir Karton}
    \vspace{1em}

    $^{\dagger}$These authors contributed equally and are ordered randomly.
    
    {
    \itshape
    $^1$Microsoft Research, AI for Science \\
    $^2$Microsoft Research, Accelerator \\
    $^3$School of Science and Technology, University of New England, Australia
    }

    $^*$\texttt{\small sehlert@microsoft.com}, \texttt{\small jan.hermann@microsoft.com}, \texttt{\small amir.karton@une.edu.au}

    \vspace{1.5em}

  \begin{minipage}{0.85\linewidth}
    \small
    \paragraph{Abstract}
Accurate thermochemical data with sub-chemical accuracy (within 1\,kcal\,mol$^{-1}$ of the empirical ground truth) are essential for advancing computational chemistry methods.
However, existing datasets that reach this level of accuracy remain limited in size or scope.
This hinders the development of data-driven methods with predictive accuracy across the broad chemical space of closed-shell, neutral molecules.
Here we present Microsoft Research Accurate Chemistry Collection (MSR-ACC) and its first release, MSR-ACC/TAE25, comprising \molcount{} total atomization energies at the CCSD(T)/CBS level obtained with the W1-F12 thermochemical protocol.
The dataset is constructed to exhaustively cover the chemical space of closed-shell, charge-neutral, covalently bound equilibrium molecular structures containing up to 5 non-hydrogen atoms drawn from elements up to argon and lacking significant multireference character.
The dataset and its canonical train and validation splits are openly available on Zenodo in the QCSchema format under the CDLA Permissive 2.0 license.
This first release of MSR-ACC enables data-driven approaches for developing predictive computational chemistry methods with unprecedented accuracy and scope.

  \end{minipage}
  \end{center}
\vspace{1em}

\section*{Background \& summary}

The enthalpy of formation is the most fundamental thermodynamic property of a molecule, from which many other thermodynamic properties are derived, for example, reaction and combustion enthalpies, bond dissociation energies, Gibbs-free energies of formation, and equilibrium constants.
The enthalpy of formation---i.e., the energy change associated with forming a molecule from its constituent elements in their standard states---is cognate to the total atomization energy~(TAE), which is the energy required to break down a molecule into its constituent atoms in the gas phase.
These two thermochemical properties are readily interconverted via the atomic enthalpies of formation at \qty{0}{\kelvin}.
From a theoretical perspective, the TAE provides a fundamental reference for determining rigorous error bars for electronic structure theories.\cite{Irikura2014,G3_05,ATcT_1,ATcT_2,G3_99,G2-1}
The reason for this is not only because the TAE is the most fundamental thermodynamic quantity, but also because it is one of the most challenging thermodynamic properties for electronic structure methods.
It is well established that thermochemical transformations become increasingly more challenging for approximate electronic structure methods as they conserve less of the chemical environments between reactants and products.\cite{Wheeler2009,karton_benchmark_2009,karton_w411_2011,Ramabhadran2011,Wheeler2012,Ramabhadran2012,Ramabhadran2014,Yu2014,Karton2016,Chan2021,Karton2023,Hehre1970,Radom1971}
For example, the performance of any approximate electronic structure method is expected to improve along the reaction hierarchy:
\begingroup
\addtolength\leftmargini{-0.1in}
\begin{quote}
atomization $\rightarrow$ isogyric $\rightarrow$ isodesmic $\rightarrow$ hypohomodesmotic $\rightarrow$ homodesmotic $\rightarrow$ hyperhomodesmotic
\end{quote}
\endgroup
This improvement is due to progressively more systematic error cancellation between reactants and products occurring along this series.\cite{Wheeler2009,Wheeler2012,Karton2016}
Thus, TAEs, which involve the complete dissociation of a molecule into its constituent atoms, represent the most extreme scenario where there is no conservation of the chemical environments between reactants and products.
It follows that the performance of electronic structure methods against TAEs provides upper-bound errors, and the performance for less challenging chemical systems and properties should improve along this series.

Over the past two decades, a number of high-level theoretical datasets for TAEs and heats of formation labeled with ab initio wavefunction methods have been developed.
They include the W4 series of datasets 
of TAEs of small compounds containing H, B, C, N, O, F, Al, Si, P, S, and Cl \cite{W4-08,karton_w411_2011,karton_w417_2017} calculated at the full configuration interaction (FCI) complete-basis-set (CBS) limit with the W4 composite wavefunction method (or variants thereof).\cite{karton_w4_2006,Karton2016wires,karton_quantum_2022,Karton2023CCC}
The prohibitive computational cost of these calculations limits the size and number of the systems, but offers the possibility to include also molecules with strong multireference character,
since the W4 protocol provides confident sub-chemical accuracy even for such systems.\cite{karton_w4_2006,W4.4,karton_w411_2011,Karton2016wires,karton_quantum_2022,Karton2023CCC}
Moving to computationally more economical coupled-cluster methods, namely with single, double, and quasiperturbative triple excitations (CCSD(T)),\cite{ccsdt} allows for the construction of datasets with many more molecules.
In a tour de force study, \citet{Narayanan2019} calculated G4(MP2) energies for 133k organic species (composed of H, C, N, O, and F) with up to 9 non-hydrogen atoms in the GDB-9 dataset (GDB9-G4MP2).\cite{ramakrishnan_quantum_2014}
G4(MP2) is a computationally efficient composite wavefunction method \cite{g4mp2,gn_review},
which includes an empirical `higher-level correction' (HLC) term
to account for efficiency-driven compromises in the theoretical model.
As a result, G4(MP2) is applicable even to large organic systems, such as large polycyclic aromatic hydrocarbons and fullerenes \cite{Karton2023,C60_1,C60_2,C40_1,C40_2}, but
it may not achieve chemical accuracy
for nonstandard bonding situations and for systems that are not well-represented in the experimental G3/05\cite{G3_05} training set.\cite{ClFn,Sn,Noblegas,PSn,CBn,Cn}
Composite wavefunction methods such as W$n$,\cite{w1} W$n$-F12,\cite{w1-f12} and W$n$X\cite{w1x} ($n$ = 1, 2) overcome these limitations by extrapolating the Hartree--Fock, CCSD, and (T) components separately to the CBS limit and also including a core-valence (CV) correction.
They were recently used to construct a number of datasets of heats of formation and TAEs \cite{Chan2022jpa,Chan2025,Karton2025cpl}, but these are mostly limited to organic systems such as those already covered by the GDB9-G4MP2 dataset \cite{Narayanan2019}.
Departing from the coupled cluster family of methods, \citet{khan2024quantummechanicaldataset836k}
developed the VQM24 dataset of chemically diverse molecules from the first three periods of the periodic table, and labeled the subset of the 10k smallest molecules with up to 4 non-hydrogen atoms with single-determinant diffusion quantum Monte Carlo (DMC).
However, DMC with a single-determinant Slater--Jastrow ansatz is insufficiently accurate for atomization energies \cite{MoralesJCTC12}, with mean absolute error on the G1 dataset \cite{g1set} of $\sim$3\,kcal\,mol$^{-1}$, and can only be made more accurate through much more expensive multideterminant ansatzes.

A significant gap in the literature is thus the absence of a chemical dataset covering a broad and diverse set of closed-shell, neutral TAEs for organic and inorganic compounds at sub-chemical accuracy.
Here, we partially fill that gap by introducing Microsoft Research Accurate Chemistry Collection (MSR-ACC) and its first dataset, designated MSR-ACC/TAE25 \cite{zenodo_dataset},
comprising \molcount{} total atomization energies at the CCSD(T)/CBS level obtained with the W1-F12 thermochemical protocol.
The dataset is constructed to exhaustively cover the chemical space of closed-shell, charge-neutral, covalently bound equilibrium molecular structures containing up to 5 non-hydrogen atoms drawn from elements up to argon and lacking significant multireference character.
Rare-gas elements were a priori excluded as they do not form covalent bonds, while all other elements up to argon (i.e., H, Li, Be, B, C, N, O, F, Na, Mg, Al, Si, P, S, and Cl) are featured prominently.
The molecular structures were obtained by exhaustive sampling of all possible molecular graphs subject in the most general case only to the constraint on a maximum valency of each element.
The resulting graphs were then converted to equilibrium 3D structures by a cascade of structure optimization steps with increasingly accurate methods.
All molecular structures that dissociated into covalently disconnected fragments, for which the triplet state was found lower in energy than a singlet state, and for which after the W1-F12 labeling the multireference character was found too strong, were discarded.
Next to the W1-F12 reference values, the dataset also includes a number of auxiliary values such as atomization energies with DFT or quantities used for filtering that serve for technical validation.

MSR-ACC/TAE25 opens the way for developing and evaluating generally applicable machine-learning, density-functional theory, and semi-empirical methods, and has already been used for training the first exchange--correlation functional to reach chemical accuracy for atomization energies \cite{luise_accurate_2025}. In contrast to smaller or more specialized TAE databases, MSR-ACC/TAE25 provides a large and chemically diverse test set for identifying systematic errors and validating approximate electronic structure methods, for example, in describing bond energies across a wide range of challenging s- and p-block compounds. The size, diversity, and accuracy of the dataset make it useful not only for developing deep-learning DFT methods,\cite{luise_accurate_2025} but also for training and validating models such as graph neural networks, critically testing their ability to generalize beyond the typical organic chemistry focus of datasets like GDB-9. Finally, the dataset can be filtered to create highly specific benchmarks to answer practical questions, such as selecting the most accurate functional for a particular type of chemical system (e.g., phosphorus sulfide compounds).\cite{PSn} We envision continuing to expand MSR-ACC in the future with datasets of quality equal to or exceeding that of the dataset presented in this work.

\section*{Methods}
\subsection*{Structure generation}

\begin{figure}[tbp]
    \centering
    \begin{tikzpicture}
    \node[below right, inner sep=0pt] at (0.0,0) {\includegraphics[width=0.8\linewidth]{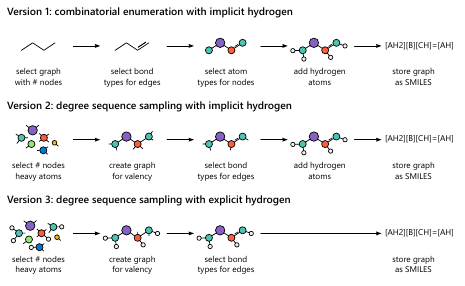}};
    \node[below right] at (-0.5,-0.1) {\bfseries\sffamily\small a};
    \node[below right, inner sep=0pt] at (0.0,-8.5) {\includegraphics[width=0.8\linewidth]{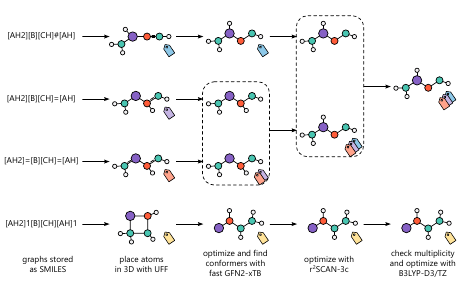}};
    \node[below right] at (-0.5,-8.8) {\bfseries\sffamily\small b};
    \end{tikzpicture}
    
    \caption{
    End-to-end molecular structure generation.
    (\textbf a) Three methods used for generating new molecular graphs (as SMILES strings).
(1) Enumerate all possible graphs, bond types, and atom types before saturating with hydrogen atoms.
(2) Sample nonhydrogen atoms, a graph and bond types to satisfy maximum allowed valencies, and saturate with hydrogen atoms.
(3) Sample nonhydrogen atoms with explicit hydrogens and then a graph and bond types to satisfy maximum allowed valencies.
    (\textbf b) From molecular graphs to final B3LYP-D3(BJ)/def2-TZVPP optimized structures.
Atoms are placed using UFF initially and then optimized using GFN2-xTB which also samples the conformational space.
Finally, the geometry is optimized using a meta-GGA (r\textsuperscript{2}SCAN-3c) and a hybrid functional (B3LYP).
Throughout, duplicates based on reconstructed molecular graph and matching TAE are merged (indicated by label symbols).
Spin states are checked with B3LYP.
    }
    \label{fig:structure-generation}
\end{figure}

The molecular geometries were generated to cover the chemical space for elements of the first three periods of the periodic table (H--Ar) excluding the rare-gas atoms.
The structure generation has two phases: molecular graph generation and conversion of molecular graphs to molecular 3D structures.

\subparagraph{Molecular graphs}
We constrain the number of electrons to be even (closed-shell molecules) and the total molecular charge to be zero.
Compared to what is common in organic chemistry, we use much looser criteria on which molecular graphs are allowed and put the only constraint on the \emph{maximum} valency of each element: 1 for Li and Na; 2 for Be, O, and Mg; 4 for B, C, N, and Cl; and 6 for every other element.
The molecular graphs were generated using different approaches to maximize the diversity of all possible bonding situations (\cref{fig:structure-generation}a).
First, we use a brute-force enumeration of all possible molecular graphs based on bare template graphs for a given number of non-hydrogen atoms (Version 1).
While this approach has combinatorial complexity, it allows us to exhaustively enumerate all possible graphs up to four non-hydrogen atoms.
Second, a degree sequence sampling approach is used where candidate graphs are considered from a list of atoms with a degree up to the respective open valency.
The resulting bare graphs were dressed with different bond types (single, double, triple), respecting the maximum valency for each atom.
The degree sequence sampling was performed once with implicitly added hydrogen atoms (Version 2) and once with explicit hydrogen atoms in the degree sampling step (Version 3).
A pseudocode for this molecular graph generation algorithm is released alongside the dataset \cite{zenodo_dataset}.
Third, an autoregressive model based on a GPT-2 transformer architecture was used for predicting new molecular graphs.
The model was trained on all SMILES strings with 5 non-hydrogen atoms generated by the graph sampling described above that could be successfully converted and optimized at B3LYP-D3(BJ)/def2-TZVPP level of theory ($\sim$6M).
The SMILES generative model was then sampled for $\sim$1.5M novel molecular graphs which were not part of the training data with a 85\% novelty success rate.
After 3D structure generation and optimization (see below), this led to about $\sim$20\% of the resulting molecular structures with 5 non-hydrogen atoms originating from the GPT-2 SMILES model.
All three approaches taken together, this exhaustive generation of molecules can be considered a bottom-up counterpart to the top-down fragment extraction of \citet{huang_quantum_2020}.

\subparagraph{3D structures}
The initial 3D structures were generated with UFF\cite{rappe_uff_1992} using OpenBabel\cite{oboyle_open_2011} to convert SMILES to 3D geometries (\cref{fig:structure-generation}a).
The resulting structure guess was optimized using GFN2-xTB\cite{bannwarth_gfn2xtban_2019} with the \textit{xtb}\cite{bannwarth_extended_2021} program package and reoptimized using r\textsuperscript{2}SCAN-3c\cite{grimme_r2scan3c_2021} and refined using B3LYP-D3(BJ)\cite{becke_densityfunctional_1993,stephens_initio_1994}/def2-TZVPP (as prescribed by the W1 protocol) with the Orca\cite{neese_software_2022} program package.
In each optimization step, structures were compared with a global index to find duplicates based on the molecular graph\cite{spicher_robust_2020} and the total atomization energy at the respective level of theory.
Importantly, all the species in the dataset were verified to be equilibrium structures by confirming they have all-real harmonic frequencies at the same level of theory.

To ensure that the dataset contains only molecules in the electronic ground state, we filter out structures for which the singlet--triplet gap $\mathrm S_0$--$\mathrm T_1$ is negative according to B3LYP/def2-TZVP---this filters out $\sim$5\% of all molecules we find.
To select the structures which can be safely treated with CCSD(T)-based composite wavefunction protocols, the fraction of TAE accounted for by the parenthetical connected triple excitations (\%TAE[(T)]) is computed at the CCSD(T)/6-31G(d) level of theory, all structures with more than 6\% \%TAE[(T)] were not considered for labeling.
This led to exclusion of additional $\sim$5\% of the generated structures.
We later verified that the \%TAE[(T)] diagnostic taken from W1-F12 theory also does not exceed 6\%.
Finally, we filter out all structures that have dissociated to more than one covalently bound fragment.
We determine the molecular graph through a combination of the bond model of GFN-FF \cite{spicher_robust_2020} and a simple heuristic based on a sum of covalent radii.

For all structures we evaluate a range of electronic structure methods, including semi-empirical methods GFN1-xTB\cite{grimme_robust_2017} and GFN2-xTB,\cite{bannwarth_extended_2021} composite electronic structure methods PBEh-3c,\cite{grimme_consistent_2015} B97-3c,\cite{brandenburg_b973c_2018} r\textsuperscript{2}SCAN-3c,\cite{grimme_r2scan3c_2021} (meta-)GGAs B97M-V\cite{mardirossian_mapping_2015} and r\textsuperscript{2}SCAN-D3(BJ),\cite{ehlert_r2scand4_2021} (range-separated) hybrid functionals B3LYP-D3(BJ),\cite{becke_densityfunctional_1993,stephens_initio_1994} M06-2X,\cite{zhao_m06_2008} \(\omega\)B97X-V,\cite{mardirossian_wb97xv_2014} and \(\omega\)B97M-V,\cite{mardirossian_wb97mv_2016} and double hybrid functionals  revDSD-PBEP86-D3(BJ)\cite{santra_minimally_2019} and \(\omega\)B97X-2.\cite{chai_longrange_2009}

\subparagraph{Subsampling}
After generating $\sim$1M unique molecular 3D structures with up to 5 non-hydrogen atoms, we subsample the structures for labeling with the high-accuracy W1-F12 protocol (see below) to maximize the chemical diversity while minimizing the total computational cost of the labeling.
We subsample along three axes: (i) number of non-hydrogen atoms, (ii) presence of s-block elements, and (iii) presence of elements from the 3rd period of the periodic table.
For each bracket, we have estimated the computational cost of W1-F12 labeling based on preliminary calculations on a small subset of molecules and then allocated a computational budget for each bracket, attempting to maximize the eventual training signal of the whole dataset at a fixed total computational cost.

\subsection*{Labeling with the composite W1-F12 protocol}

Using the fully optimized B3LYP-D3(BJ)/def2-TZVPP structures, we obtain the nonrelativistic, all-electron CCSD(T)/CBS TAE using the high-level W1-F12 composite wavefunction protocol.\cite{w1-f12} W1-F12 theory is an explicitly correlated\cite{F12_1,F12_2,F12_3} version of the original W1 theory.\cite{w1} The computational details of W1-F12 theory have been specified and rationalized in great detail elsewhere.\cite{w1-f12} In brief, the Hartree--Fock (HF) energy is extrapolated to the complete basis-set limit from the cc-pVDZ-F12 and cc-pVTZ-F12 basis sets,\cite{peterson_systematically_2008} using the $E(L) = E_\infty + A/L^\alpha$ two-point extrapolation formula, with $\alpha$ = 5.
The complementary auxiliary basis set (CABS) singles correction is included in the Hartree--Fock (HF) energy.\cite{CABS_1,CABS_2,CABS_3}
The valence CCSD-F12 correlation energy (denoted by $\Delta$CCSD hereinafter) is extrapolated from the same basis sets, using an optimal extrapolation exponent of $\alpha$ = 3.67 for elements up to argon. Optimal values for the geminal Slater exponents were taken from \citet{peterson_systematically_2008} and \citet{Hill2010}, and the resolution of the identity (RI) approximation was applied using the OptRI auxiliary basis sets.\cite{Yousaf2008} The diagonal, fixed-amplitude 3C(FIX) ansatz and the CCSD-F12b approximation were used in these calculations.\cite{CABS_2,CABS_3} The valence parenthetical connected triple excitations contribution (denoted by $\Delta$(T)) is obtained from standard CCSD(T) calculations, as is the case in W1w theory.\cite{w1,karton_w4_2006} Specifically, the $\Delta$(T) component is extrapolated from the jul-cc-pV(D+d)Z and jul-cc-pV(T+d)Z basis sets using the above two-point extrapolation formula with $\alpha = 3.22$.\cite{dunning1,dunning2,dunning3,papajak_perspectives_2011} The CCSD inner-shell contribution is calculated with the core-valence weighted correlation-consistent cc-pwCVTZ basis set of Peterson and Dunning,\cite{dunning4} while the (T) inner-shell contribution is calculated with the cc-pwCVTZ basis set without the f functions. 
Since we are primarily interested in assessing the performance of nonrelativistic computational chemistry methods on the electronic potential energy surface, the scalar relativistic, spin-orbit, and diagonal Born-Oppenheimer corrections are not considered in the present work. All the calculations involved in W1-F12 theory were carried out with the Molpro 2024.1 program suite.\cite{Werner2012,Werner2020,MOLPRO_full}

We have labeled all found molecules with up to 4 non-hydrogen atoms, and a smaller sample of all found molecules with 5 non-hydrogen atoms.

\section*{Data records}

The data records for \molcount{} molecules of MSR-ACC/TAE25 are available in Zenodo under the CDLA Permissive 2.0 license \cite{zenodo_dataset}.
To aid machine-learning applications, we also release training and validation subsets of the full dataset, which constitute canonical 99\% and 1\% splits, respectively, of MSR-ACC/TAE25 after removing the overlap with the popular W4-17 and GMTKN55 benchmark sets.
The overlap was constructed such that it contains any atomization reaction where the reconstructed molecular graphs of the molecules match.

Molecules are formatted using the QCSchema standard, including the Cartesian coordinates in bohr (\verb+geometry+ field), element symbols (\verb+symbols+ field), total charge (\verb+molecular_charge+ field) and spin state (\texttt{molecular\_\allowbreak multiplicity} field).
The \verb+extras+ field of each molecule contains the W1-F12 atomization energy labels (\verb+tae@w1-f12+) as well as additional information including molecular graphs, DFT atomization energies, singlet--triplet gaps, and W1-F12 energy components described in \cref{tbl:extras}.

\begin{table}[t]
    \centering
    \caption{Labels included in the \texttt{extras} field of the data records.}
    \label{tbl:extras}
    \small
    \begin{tabular}{*{3}{c}}
    \toprule
    Key & Description & Unit \\
    \midrule
    \verb+graph:all+ & Connectivity of all atoms & list of indices (0-based)\\
    \verb+graph:non-h+ & Connectivity of all non-hydrogen atoms & list of indices (0-based) \\
    \verb+tae@{method}+ & Total atomization energy & hartree \\
    \verb+tae:frac[(T)]@{method}+ & \%TAE[(T)] as a fraction & 1 \\
    \verb+singlet-triplet-gap-s0-t1@{method}+ & Singlet--triplet gap S$_0$--T$_1$ & hartree \\
    \verb+tae[{component}]@w1-f12+ & Component of the W1-F12 TAE & hartree \\
    \bottomrule
    \end{tabular}
\end{table}

\section*{Technical validation}

\begin{figure}[t]
    \centering
    \includegraphics{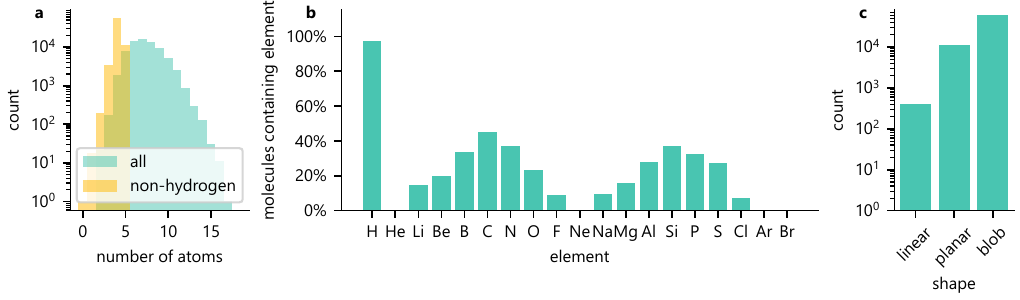}
    \caption{Elemental and structural distributions.
    (\textbf a) Molecular size as atom count.
    (\textbf b) Elemental occurrence. Hydrogen is present in 97.6\% of molecules.
    (\textbf c) General 3D shape of the molecules.
    }
    \label{fig:basic-dist}
\end{figure}

\Cref{fig:basic-dist} gives an overview of basic structural distributions over MSR-ACC/TAE25.
The molecular sizes range from zero (H$_2$) to five non-hydrogen atoms and from 2 to 17 (isopentane) atoms including hydrogen.
Defining organic systems as molecules containing at least one carbon atom, there are 45.1\% organic systems and 54.9\% inorganic systems.
Thus, similarly to the much smaller and chemically less diverse W4-17 dataset, there is a similar representation of organic and inorganic systems.
The most represented elements in the dataset are:
\begin{quote}
C > N and Si > B and P > S and Al > O > Be > Mg > Li > Na > F > Cl
\end{quote}
Importantly, about 75\% of the systems are mixed second/third-period species, 17\% are pure second-period systems, and about 8\% are pure third-period systems.
The geometry of the molecules spans linear (0.6\%), planar (15.2\%) and general 3D structures (84.3\%).

\begin{figure}[tb]
    \centering%
    \includegraphics{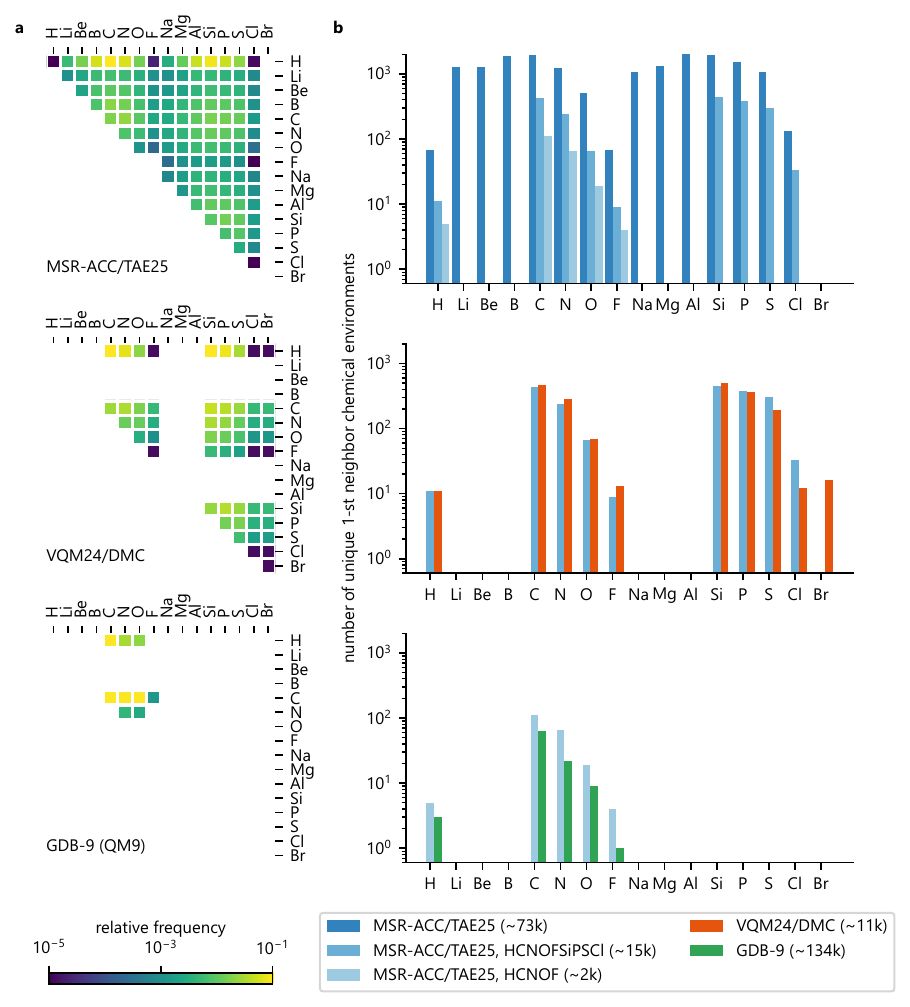}
    \caption{Reconstructed molecular graphs compared across three datasets: MSR-ACC/TAE25, VQM24/DMC, and GDB-9 (also known as QM9).
    Graphs are reconstructed from 3D structures using a bond model from GFN-FF and a simple heuristic based on a sum of covalent radii.
    (\textbf a) Distribution of element pairs in bonds.
    (\textbf b) Unique 1-st neighbor environments per element.
    Subsets of MSR-ACC/TAE25 constrained to elemental subspaces are also shown.
    }
    \label{fig:bond-graph}
\end{figure}

The elemental composition of the molecules in MSR-ACC gives rise to a diverse bond types, with many molecules involving nontraditional bonding situations.
Overall, there are 287k bonds between non-hydrogen atoms in MSR-ACC/TAE25 according to our bond heuristic, covering all possible combinations of elements in the first two rows except for F--F.
The occurrence of element pairs involved in bonding follows the pattern that could be expected from typical valencies of individual elements (\Cref{fig:bond-graph}a).
Notable outliers in occurrence are pairs of typically monovalent elements: H--H (only once in H$_2$), H--halogen (20), Cl--halogen (7), and F--F (none, as F$_2$ is associated with \%TAE[(T)] > 6\%).
This can be contrasted to two other large atomization energy datasets labeled with ab initio wavefunction methods.
VQM24/DMC follows the same bond distributions with non-hydrogen elements from groups 1, 2, and 13 (Li, Na, Be, Mg, B, Al) omitted and Br added (following the same pattern as other halogens).
GDB-9 is severely restricted in diversity by its focus on organic drug-like molecules, resulting in only C--C, C--N, C--O, C--F, N--N, and N--O non-hydrogen bonds, with the latter three occurring with low relative frequency.

A different metric of chemical diversity is obtained by counting unique 1st-neighbor environments in the molecular graphs for each element (\Cref{fig:bond-graph}b).
In the full dataset, all elements except for H and halogens occur in $\sim$1000 unique chemical environments, while H and halogens occur in $\sim$100 unique chemical environments.
Hydrogen gains this diversity through various two-electron three-centric bonds with one atom from groups 1, 2, and 13, and some other non-hydrogen element.
When restricting to molecules with no non-hydrogen atom from groups 1, 2, and 13, the number of unique chemical environments drops by half an order of magnitude, and its distribution over elements is remarkably similar to VQM24/DMC.
This indirectly verifies that the structure generation for both datasets exhaustively covers all possible local chemical environments in the respective elemental spaces.
When further restricting to the HCNOF subset, we find that the number of unique environments in MSR-ACC is still 2--3 times larger than in GDB-9, owing to covering more than just `drug-like' organic chemistry.

Yet another validation of the coverage of the chemical space is comparison of the overlap between MSR-ACC/TAE25 and the closed-shell single-referential subset of W4-17 up to 4 non-hydrogen atoms, which we expect to be covered exhaustively.
We find that only two molecules from this W4-17 subset are missing in MSR-ACC/TAE25 for fundamental chemical reasons: diborane, B$_2$H$_6$, which due to its uniqueness cannot be constructed by our molecular graph generation, and cyclobutadiene, C$_4$H$_4$, likely due to its instability, since we did find two dozen of its other isomers.
For three other missing molecules (hydrogen peroxide, H$_2$O$_2$, acetaldehyde, CH$_3$CHO, and hydrazine, N$_2$H$_4$), our structure generation yielded the corresponding molecular graph, but the subsequent 3D structure optimization yielded a saddle point rather than an equilibrium structure, so they were filtered out.

\begin{figure}[tb]
    \centering
    \includegraphics{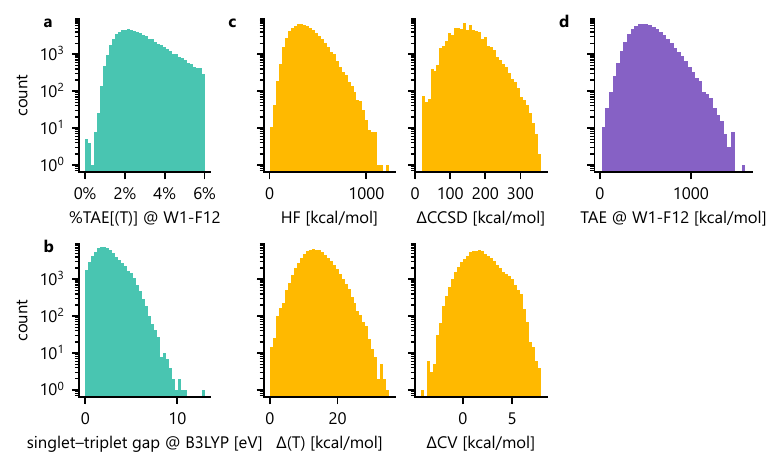}
    \caption{Energetic distributions.
    (\textbf a) Multireference diagnostic \%TAE[(T)].
    (\textbf b) Vertical singlet-triplet gaps S$_0$--T$_1$ from B3LYP.
    (\textbf c) All four components of the W1-F12 total atomization energy.
    (\textbf d) Reference W1-F12 total atomization energy.
    }
    \label{fig:basic-dist-2}
\end{figure}

It is well established that the CCSD(T) method cannot generally achieve chemical accuracy for systems dominated by moderate-to-severe nondynamical correlation effects.\cite{karton_w4_2006,karton_quantum_2022,Karton2023CCC,Karton2016wires,karton_w411_2011,karton_w417_2017,W4.4,Karton2018,Feller2014,Peterson2012,Feller2008,Feller2007}
It is therefore essential to estimate the contributions to the TAEs from post-CCSD(T) excitations.
A highly successful a priori predictor for such contributions is the so-called \%TAE[(T)] diagnostic,\cite{karton_w4_2006,Karton2016wires,karton_w411_2011} which is defined as the percentage of the TAE accounted for by parenthetical connected triple excitations:
$$
\mathrm{\%TAE}[\mathrm{(T)}] = \frac{\mathrm{TAE}[\mathrm{CCSD(T)}] - \mathrm{TAE}[\mathrm{CCSD}]}{\mathrm{TAE}[\mathrm{CCSD(T)}]}
$$
\noindent where TAE[CCSD] and TAE[CCSD(T)] are the TAEs calculated at the CCSD and CCSD(T) levels.
This simple multireference diagnostic is particularly well suited for eliminating systems with significant post-CCSD(T) contributions to the TAEs.
It has been suggested that \%TAE[(T)] values up to $\sim$5\% indicate that post-CCSD(T) contributions to the TAEs should not exceed $\sim$0.5\,kcal\,mol$^{-1}$.\cite{Karton2016wires}
In the present work, we have chosen to exclude all systems with $\mathrm{\%TAE}[\mathrm{(T)}]> 6\%$.
The distribution of \%TAE[(T)] (\cref{fig:basic-dist-2}a) therefore ends sharply at 6\%, while peaking at $\sim$2\% and tapering off towards 0\%.
The second major filtering criterion is positive vertical singlet--triplet S$_0$--T$_1$ gap, in order to have only molecules in their electronic ground states.
The distribution of the S--T gap (\cref{fig:basic-dist-2}b) therefore ends sharply at 0\,eV, peaks at $\sim$2\,eV, and tapers off to $\sim$10\,eV.

\cref{fig:basic-dist-2}c gives an overview of the HF, $\Delta$CCSD, $\Delta$(T), and $\Delta$CV contributions to the W1-F12 TAEs.
The HF component spreads over a wide range from 1.3 to 1238.8\,kcal\,mol$^{-1}$.
We note, however, that small HF TAEs below 32.5\,kcal\,mol$^{-1}$ are obtained for only 12 systems, nearly all of which consist of Li bonding with Na, Be, or Mg.
The overall all-electron CCSD(T)/CBS TAEs for these 12 systems are also relatively small and range between 20.1 and 62.2\,kcal\,mol$^{-1}$.
HF TAEs above 1000\,kcal\,mol$^{-1}$ are obtained for 38 systems, nearly all of which contain 3--4 carbon.
We note that there is no apparent correlation between the HF component and the number of non-hydrogen atoms, however, the squared correlation coefficient between the total number of atoms and the HF component is $R^2 = 0.757$.

The $\Delta$CCSD component spreads over a wide range of 340\,kcal\,mol$^{-1}$ from 18.4 to 358\,kcal\,mol$^{-1}$.
The largest 40 $\Delta$CCSD contributions range between 327 and 358.1\,kcal\,mol$^{-1}$.
Most of these systems are nitrogen-rich species.
The systems with the smallest $\Delta$CCSD contributions are dominated by beryllium compounds.
We note that overall, for the \molcount{} systems in the dataset, there is a noticeable but weak statistical correlation between $\Delta$CCSD component and the number of valence electrons, with $R^2$ = 0.613.

The $\Delta$(T) component varies between 0.0 and 35.3\,kcal\,mol$^{-1}$.
The lowest $\Delta$(T) contributions are obtained, as expected, for lithium and sodium systems.
Excluding the systems with only two valence electrons (H$_2$, HLi, HNa, LiNa) for which the $\Delta$(T) contribution is zero by definition, the lowest 40 $\Delta$(T) contributions range between 0.0 and 1.8\,kcal\,mol$^{-1}$.
Of these 40 systems, 23 contain either Na or Li, or both, where in most cases there are multiple Li and Na atoms in the same molecule.
Similarly to the $\Delta$CCSD component, the largest $\Delta$(T) contributions are obtained for nitrogen-rich species.

The $\Delta$CV contribution can be either negative or positive, meaning that $\Delta$CV correlation is more or less stabilizing, respectively, in the constituent atoms than in the molecule.
Overall, the $\Delta$CV contributions range from $-$4.2 to +7.9\,kcal\,mol$^{-1}$, where 12\% of the $\Delta$CV values are negative and 88\% are positive.
In line with observations\cite{Karton2016wires,karton_quantum_2022,Karton2023CCC} on the W4-17 dataset,\cite{karton_w417_2017} 87.5\% of the systems with a negative CV correction contain Al, Si, or both atoms.
The resulting TAEs range from 20.1 (LiNa) to 1560\,kcal\,mol$^{-1}$ (isopentane).

\begin{figure}[tb]
    \centering
    \includegraphics{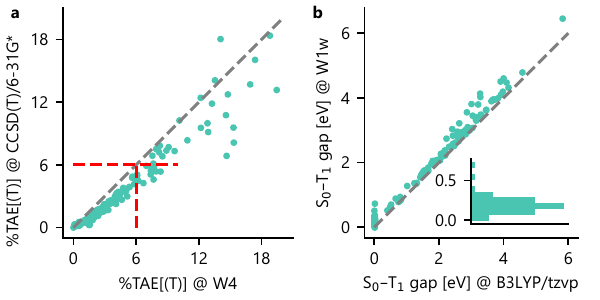}
    \caption{Validation of filtering criteria.
    (\textbf a) Multireference descriptor \%TAE[(T)] evaluated on the W4-17 dataset in the small 6-31G(d) basis almost always underestimates the true CBS value (from W4), leading to false negatives but no false positives of multireference character.
    (\textbf b) Singlet--triplet gaps are only overestimated by B3LYP, never underestimated.
    A random sample of molecules with the B3LYP gap close to zero (inset) has the CCSD(T) gap (from W1w) ranging from 0.0 to 0.4\,eV.
    }
    \label{fig:w4-17-pt-tae}
\end{figure}

\begin{figure}[!bt]
    \centering
    \includegraphics[width=0.96\linewidth]{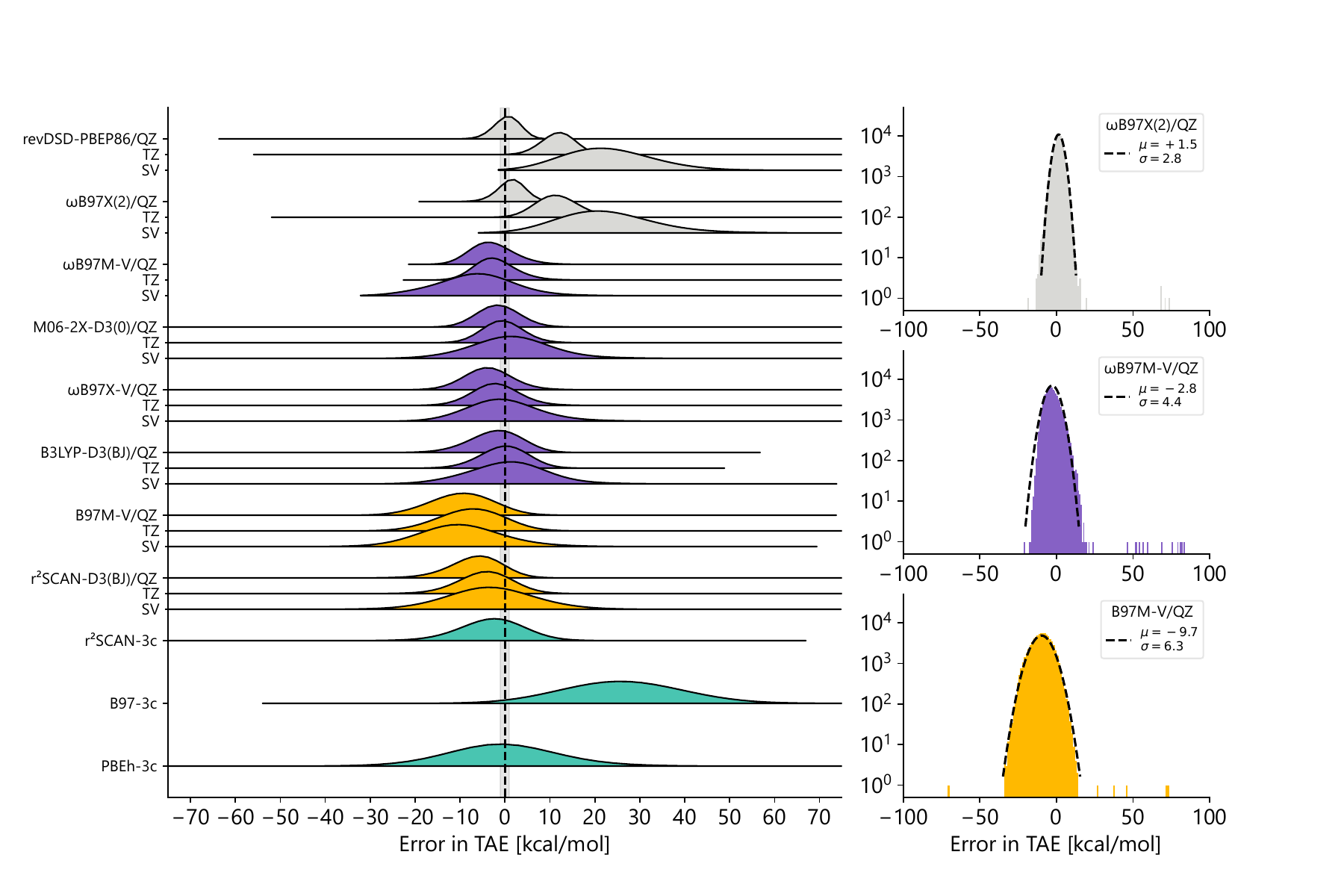}
    \caption{Error distribution of common exchange--correlation (XC) functionals with respect to W1-F12 reference values.
    The error distribution of each functional is shown in def2-QZVP, def2-TZVP, and def2-SVP from top to bottom, except for composite DFT methods.
    The error range of \(\pm1\)\,kcal\,mol$^{-1}$ is highlighted as light gray area.
    Jacob's ladder rungs are denoted with color.
    For selected functionals the distribution is replotted in log scale with a fitted normal distribution.
    }
    \label{fig:baseline-tae}
\end{figure}

We introduced two primary filtering criteria when constructing MSR-ACC.
First, \%TAE[(T)] computed in the small 6-31G(d) basis is used to identify multireference molecules to be filtered out before labeling them with the much more expensive W1-F12 calculation.
We have verified on the W4-17 dataset that this almost never leads to excluding molecules that are in fact single-referential, as the small-basis version of the metric strictly underestimates it (\cref{fig:w4-17-pt-tae}a).
After calculating the W1-F12 TAEs, we again removed any additional systems with a \%TAE[(T)] diagnostic above 6\%, where the \%TAE[(T)] CBS values were taken from W1-F12 theory.
It is important to note that this filtering process may result in some `false negatives' (i.e., erroneously flagging well-behaved molecules as having multireference character) but is highly unlikely to produce `false positives' (i.e., erroneously including multireference molecules).\cite{Karton2016wires} Second, singlet--triplet gap S$_0$--T$_1$ is approximated by B3LYP/def2-TZVP to exclude molecules with a triplet ground state.
We have verified on a random sample of molecules from the dataset against gaps calculated with W1w that B3LYP strictly underestimates the gap (\cref{fig:w4-17-pt-tae}b).
This leads to a small fraction of molecules being unnecessarily filtered out that in fact have a singlet ground state, but we consider this worth the saved cost of having to run extra W1-F12 triplet calculations.
This ensures that the molecules that pass this filter are reliably singlet ground-state systems.
The conservative combination of the \%TAE[(T)] diagnostic and the B3LYP singlet--triplet gap ensures the thermochemical reliability of the final MSR-ACC/TAE25 dataset.

Next to the reference W1-F12 values, the dataset contains TAEs from KS-DFT with a number of popular exchange-correlation (XC) functionals.
We found that the distribution of errors of all functionals with respect to the reference values follows the normal distribution very closely, which enables us to identify outliers confidently (\cref{fig:baseline-tae}).
This can then serve as an indirect validation of the reference values, where we ensure that there are no cases of different XC functionals agreeing among each other on the TAE but disagreeing with W1-F12.
The accuracy of the different XC functionals follows the general trends one would expect from the Jacob's ladder categorization.

The MSR-ACC/TAE25 dataset provides a large-scale, internally consistent dataset of CCSD(T)/CBS TAEs calculated with the high-level W1-F12 composite wavefunction method\cite{w1-f12}. Our technical validation focuses on ensuring that the molecules included in the dataset can be reliably treated at this level of theory. By applying the \%TAE[(T)] diagnostic and the singlet–triplet gap, we systematically filter out molecules for which the CCSD(T) method would be less reliable. It is important to note that W1-F12 theory has been extensively benchmarked against highly accurate experimental and theoretical TAEs. For example, for the 137 TAEs in the W4-11 database, which provides reference values approximating the relativistic, all-electron FCI/CBS level of theory via W4 theory,\cite{karton_w4_2006,karton_w411_2011} W1-F12 theory attains a mean absolute deviation (MAD) of 0.51 kcal\,mol$^{-1}$ and a root-mean-square deviation (RMSD) of 0.74 kcal\,mol$^{-1}$. \cite{w1-f12} These error statistics are reduced to MAD = 0.32 and RMSD = 0.65 kcal\,mol$^{-1}$ when post-CCSD(T) contributions are removed from the W4 reference values.\cite{w1-f12} It should be noted that W4 theory attains a MAD and RMSD of 0.065 and 0.085 kcal\,mol$^{-1}$, respectively, relative to highly accurate experimental TAEs obtained from the active thermochemical tables (ATcTs) thermochemical network.\cite{karton_w411_2011,Ruscic2004, Ruscic2005, Ruscic2006} While the primary goal of the present work is to provide all-electron CCSD(T)/CBS TAEs for the parameterization and validation of electronic structure methods, these reference TAEs can be directly compared with experimental values by adding scalar relativistic, spin-orbit, diagonal Born–Oppenheimer corrections (DBOCs), and zero-point vibrational energies (ZPVEs).\cite{Karton2016wires,karton_quantum_2022,Karton2023CCC}

\section*{Code availability}

OpenBabel\cite{oboyle_open_2011} and \textit{xtb}\cite{bannwarth_extended_2021} are open-source software.
Orca\cite{neese_software_2022} and Molpro\cite{Werner2012,Werner2020,MOLPRO_full} are commercial software.
The SMILES generation is specified in the Methods section and no further custom code was used.

\begingroup
\setlength\bibitemsep{0pt}
\printbibliography%
\endgroup

\section*{Author contribution}

S.E.\ designed the structure refinement pipeline.
J.H., M.S., and S.E.\ designed the graph sampling routines.
A.K.\ performed the W1-F12 calculations.
V.G.S.\ implemented the GPT-2 SMILES model.
T.V., S.L., D.P.K., C.-W.H., and G.L.\ tested the generated data.
S.E., T.V. and S.L.\ designed and built the software infrastructure.
J.H., K.T., R.v.d.B., P.G.-G., and A.K.\ supervised the research.
A.K., S.E., and J.H.\ wrote the manuscript.

\section*{Competing interests}

All authors declare employment by Microsoft.
All authors declare a filed patent application for a data generation system described in this article.

\section*{Acknowledgments}

We thank J.\ Zhou (EPFL) for bringing our attention to missing information in earlier versions of the manuscript and dataset, as well as A.\ Pedro Fontes (UCLouvain) for reporting to us inconsistencies in earlier versions of the dataset.
We acknowledge Microsoft Accelerating Foundation Models Research Program for access to Microsoft Azure cloud compute.

\end{document}